\PassOptionsToPackage{super,sort&compress}{natbib}
\documentclass[pdflatex,sn-mathphys-num]{sn-jnl}
\usepackage[utf8]{inputenc}
\usepackage{CJKutf8}
\usepackage{aas_macros}%
\usepackage{graphicx}%
\usepackage{multirow}%
\usepackage{amsmath,amssymb,amsfonts}%
\usepackage{amsthm}%
\usepackage{mathrsfs}%
\usepackage{newtxmath}
\usepackage[title]{appendix}%
\usepackage{xcolor}%
\usepackage{textcomp}%
\usepackage{manyfoot}%
\usepackage{booktabs}%
\usepackage{comment}
\usepackage{algorithm}%
\usepackage{algorithmicx}%
\usepackage{algpseudocode}%
\usepackage{listings}%
\theoremstyle{thmstyleone}%
\theoremstyle{thmstyletwo}%

\theoremstyle{thmstylethree}%

\usepackage[section]{placeins}
\usepackage{float}

\usepackage{graphicx}
\usepackage{amssymb}
\usepackage{booktabs}
\newcommand{\Msun}{M$_{\odot}$}
\newcommand{\MJup}{M$_{\mathrm{Jup}}$}
\newcommand{\RJup}{R$_{\mathrm{Jup}}$}

\newcommand{\RCD}{R$_{\mathrm{CD\mbox{-}35b}}$}
\newcommand{\CRIRES}{CRIRES$^+$}

\begin{document}
\begin{titlepage}
    \thispagestyle{empty}
    \vspace*{\fill}
        \Large\bfseries\noindent Disclaimer
        \vspace{0.6em}

        \normalsize
        \noindent This work has now been accepted for publication in \textit{Nature}. This is the paper as it existed
        before the peer review process, and the results of the accepted work, specifically which of the
        presented satellite models is favored and the exact parameters of those models, have meaningfully
        changed. Please see our final work, as published in \textit{Nature} in the coming weeks, for more
        robust results before drawing specific conclusions about this work or reporting to the public/media.
    \vspace*{\fill}
\end{titlepage}
%\title[Satellites around CD\mbox{-}35 2722 B]{Giant Exomoons? Evidence for satellite(s) orbiting substellar companion CD\mbox{-}35 2722 B}

\title[Satellites around CD\mbox{-}35 2722 B]{Planetary-Mass Exosatellite Detected Around the Substellar Companion of a Star}

\author*[1,2,3]{\fnm{Kevin} \sur{Hoy}}\email{kevin.hoy@mail.udp.cl}

\author[1,3]{Alice Zurlo}

\author[1,4]{Pablo A.~Pe{\~n}a R.}

\author[5]{Jana K\"ohler}

\author[7]{Silvano Desidera}

\author[7]{Raffaele Gratton}

\author[7,3]{Cecilia Lazzoni}

\author[6,3]{Simon Petrus}

\author[2]{Florian Rodler}

\author[2]{Jonathan Smoker}

\author[8,9]{Valentina D'Orazi}

\author[7]{Ilaria Carleo}

\author[10,1,7,3]{Ilaria Giovannini}

\affil[1]{Instituto de Estudios Astrofísicos, Facultad de Ingeniería y Ciencias, Universidad Diego Portales, Av. Ejército Libertador 441, Santiago, Chile}

\affil[2]{European Southern Observatory, Alonso de Córdova 3107, Casilla 19, Santiago 19001, Chile}

\affil[3]{Millennium Nucleus on Young Exoplanets and their Moons (YEMS)}

\affil[4]{Centro de Astrofísica y Tecnologías Afines (CATA), Casilla 36-D, Santiago, Chile}

\affil[5]{TLS Tautenburg, Sternwarte 5, 07778 Tautenburg, Germany}

\affil[6]{NASA-Goddard Space Flight Center, Greenbelt, MD 20771, USA}

\affil[7]{INAF-Osservatorio Astronomico di Padova, Vicolo Osservatorio 5, 35122 Padova, Italy}

\affil[8]{Dept. of Physics, University of Rome Tor Vergata, via della Ricerca Scientifica 1, 00133 Rome, Italy}

\affil[9]{INAF Osservatorio Astronomico di Roma, via Frascati 33, Monte Porzio Catone, 00040 Rome, Italy}

\affil[10]{Dipartimento di Fisica e Astronomia, Università degli Studi di Padova, Vicolo dell’Osservatorio 3, 35122 Padova, Italy}

%\abstract{Despite many years of searching, the sample size of confirmed moons beyond the Solar System, {\it exomoons}, remains zero. In this work, we aimed to apply the radial velocity technique, typically used to discover exoplanets around stars, to detect exomoons around the directly-imaged exoplanet CD\mbox{-}35 2722 B. Due to this planet's high luminosity and projected separation from its host star, it is possible to take high-resolution CRIRES+ (\Rlam~100,000) spectra of the planet directly, with negligible contamination from the host. By extracting the radial velocity of the planet across 20 epochs of observations, we identified a periodical oscillation which could be interpreted as a signal induced by one or more satellites. Model fits find two likely scenarios. If we force a low eccentricity, the data is best fit by a 2-exomoon model, including a large, 0.232 \MJup (3.2\% of the host mass), moon on a 168.7 day orbit, along with a smaller, 0.091 \MJup (1.2\% host mass), moon on a shorter 87.2 day orbit. Loosening the eccentricity prior finds that a single moon fits best, which is very similar to the previously mentioned large moon at 0.262 \MJup (3.6\% host mass), on a 170-day, 0.295 eccentricity orbit.}

\abstract{Despite more than 6000 exoplanets being discovered to date \cite{schneider2011}, no satellite orbiting an exoplanet, an {\it exomoon}, has ever been confidently detected. While there are some candidates, they lack clear and convincing confirmation and remain controversial \cite{lazzoni2020, oza2019, kipping_2022_k1513, kipping2022_k1708}. Beyond the innate value of discovering new types of objects in the Universe, satellites can help give key insights into planet formation mechanisms and the dynamical evolution histories of their systems. In this work, we show strong evidence for the existence of satellites orbiting the directly-imaged brown dwarf companion CD\mbox{-}35 2722 B. We have applied radial velocity analysis, the same technique used to discover the first exoplanet around a Solar-type star \cite{mayor1995}, on spectra of this brown dwarf obtained with VLT/\CRIRES. We have found what appears to be the periodic signal induced by at least one orbiting satellite. This is the first time this technique has successfully produced evidence of satellites. We produce a strong detection of a satellite candidate with a minimum mass of 0.743 Jupiter masses and an orbital period of 169 days. The best-fitting model also includes a second, closer satellite with minimum mass of 0.277 Jupiter masses and a period of 87 days, although these parameters for this smaller satellite candidate are less certain. These periods would place them very near a 2:1 mean motion resonance, a phenomenon also seen in the Galilean moons of Jupiter. The discovery of these satellites will unlock many future avenues of study, including planet formation, system dynamics, and even the search for life in the Universe \cite{heller2014}.}

%\keywords{planetary systems, planets and satellites: detection, exomoons, radial velocities}



\maketitle 

\section{Main}

CD\mbox{-}35 2722 B (hereafter CD\mbox{-}35 B for brevity) is a young brown dwarf (BD) discovered via direct imaging in 2011 \cite{wahhaj2011}. At a projected separation of $\sim$2.8'', it is well-resolved from its host star. A later study partially constrained its orbit, finding that it has a high eccentricity ($>$0.9) and a long period of $\sim$5000 years, although many parameters are poorly constrained due to the short coverage of its wide orbit \cite{bowler2020}. It is a 37 \MJup\ companion to a 0.4 \Msun\ M-type star, making the mass ratio between host and companion relatively low for this system \cite{gratton2024, wahhaj2011}. It has been calculated that, given the existence of satellites orbiting CD\mbox{-}35 B, the radial velocity method would be relatively likely to find them\cite{lazzoni2022}.

%Based on the analysis of its medium-resolution spectrum, CD\mbox{-}35 B has a mass of 7.3$\pm$1.1 \MJup, a radius of 1.6$\pm$0.05 \RJup, a spectral type of L4.5, and an age of 133$\pm$20 Myr \cite{palmabifani2025}.

%Selecting this target allows us to bypass key challenges seen in similar exomoon searches. The radial velocity (RV) method of exomoon detection has previously been attempted, specifically targeting GQ Lup B \cite{horstmann2024}, HR 7672 B \cite{ruffio2023}, and the HR8799 system \cite{vanderburg2021}. Their results were inconclusive, due to high errors in every extracted RV point. These errors are likely due to the need to correct for stellar contamination in their data, and this modelling can introduce significant uncertainties to the analysis. CD\mbox{-}35 B's far separation from its host allows us to bypass the need to remove contamination, and our RVs are orders of magnitude more precise.

We conducted a VLT/\CRIRES\cite{dorn2023} program (PI: Zurlo, 0112.C-2053(A), 0114.C-2039(A)) to monitor the radial velocity (RV) variation of the target induced by possible satellites from October 2023 through January 2025. Given the projected separation of CD\mbox{-}35 B and the narrow 0.2'' slit of \CRIRES, we were able to obtain the spectra of the companion itself, without significant contamination of the primary star. We obtained 21 epochs in total, but one is discarded in this analysis due to a very low signal. Each epoch is an AB nodding sequence, giving us the ability to subtract sky contamination while also taking two separate spectra each night. The data was reduced using the official ESO pipeline recipes.

Similar observations have been attempted in the past, targeting other direct-imaging substellar objects with different instruments. GQ Lup B \cite{horstmann2024} and HR 7672 B \cite{ruffio2023} were observed with KPIC, a high-resolution spectrograph (R$\approx$35,000) built for characterizing directly-imaged planets. The HR 8799 planets were observed with Keck/OSIRIS, a medium-resolution integral-field spectrograph (R$\approx$4000) \cite{vanderburg2021}. None of those companions showed RV variation at each work's respective precisions. Compared to those works, we are able to achieve much more precise RV results due to a combination of increased spectral resolution (R$\approx$100,000) and negligible stellar contamination, with this target having better contrast and higher angular separation from the host star. Modeling the removal of stellar contamination can be a significant source of error in RV calculations, and we are able to avoid this process entirely. See Section 2.1 of the methods for a more detailed discussion of contamination characterization.

Objects orbiting such targets, having not been discovered before, have some ambiguity regarding their nomenclature. For this work, we will use the term "satellite" to refer to the body/ies we are presenting. This is meant as a neutral term simply meaning "an object orbiting a significantly larger object", in the same sense that the Moon is a satellite of Earth, and the Earth is a satellite of the Sun. More discussion of the nomenclature is present at the end of the article.

To extract the RVs of CD\mbox{-}35 B, we used \texttt{viper}\cite{zechmeister2021, kohler2025}, a template-matching RV code with built-in \CRIRES support. This code creates an empirical template of our time series spectra, with no need to employ theoretical models. We run \texttt{viper} for all 40 (20 nights, 2 nodding positions per night) spectra of CD\mbox{-}35 B, then bin the RVs of the nodding pairs to maximize precision (more details in Methods). 

\begin{figure}[t]
    \centering
    \includegraphics[width=\linewidth]{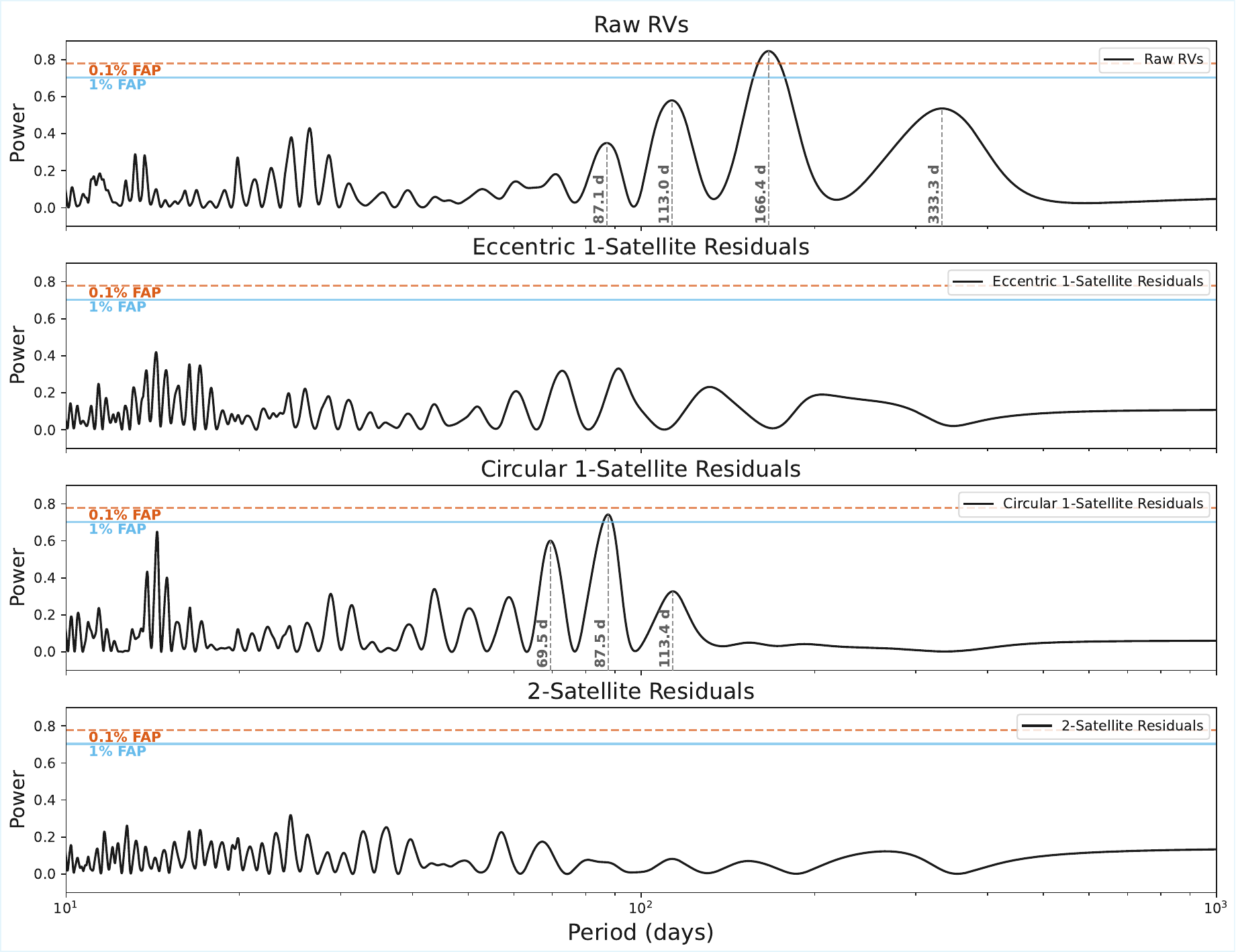}
    \caption{
        Periodogram associated with the RV values seen in Figure~\ref{fig:orbit} in the first panel. The other panels show periodograms for the residuals of the model fits.
    }
    \label{fig:periodogram}
\end{figure}

As an initial check for periodicity in our RVs, we created a Generalized Lomb-Scargle (GLS) periodogram \cite{lomb1976}. This is a function that measures how likely it is that a data set is periodic at a given period. GLS periodograms often produce false peaks aliased from artifacts in the data, and so to determine the validity of a peak, there are defined False Alarm Probability (FAP) thresholds that can be calculated based on the number of observations and their relative time sampling \cite{burt2025}. In our data, there is a peak near a 170-day periodicity that crosses the 0.1\%  FAP threshold, denoting a high degree of confidence that there is some periodic signal in the data. The GLS periodogram for our data is shown in the first panel of Fig.~\ref{fig:periodogram}.

\begin{figure}[t]
    \centering
    \includegraphics[width=\linewidth]{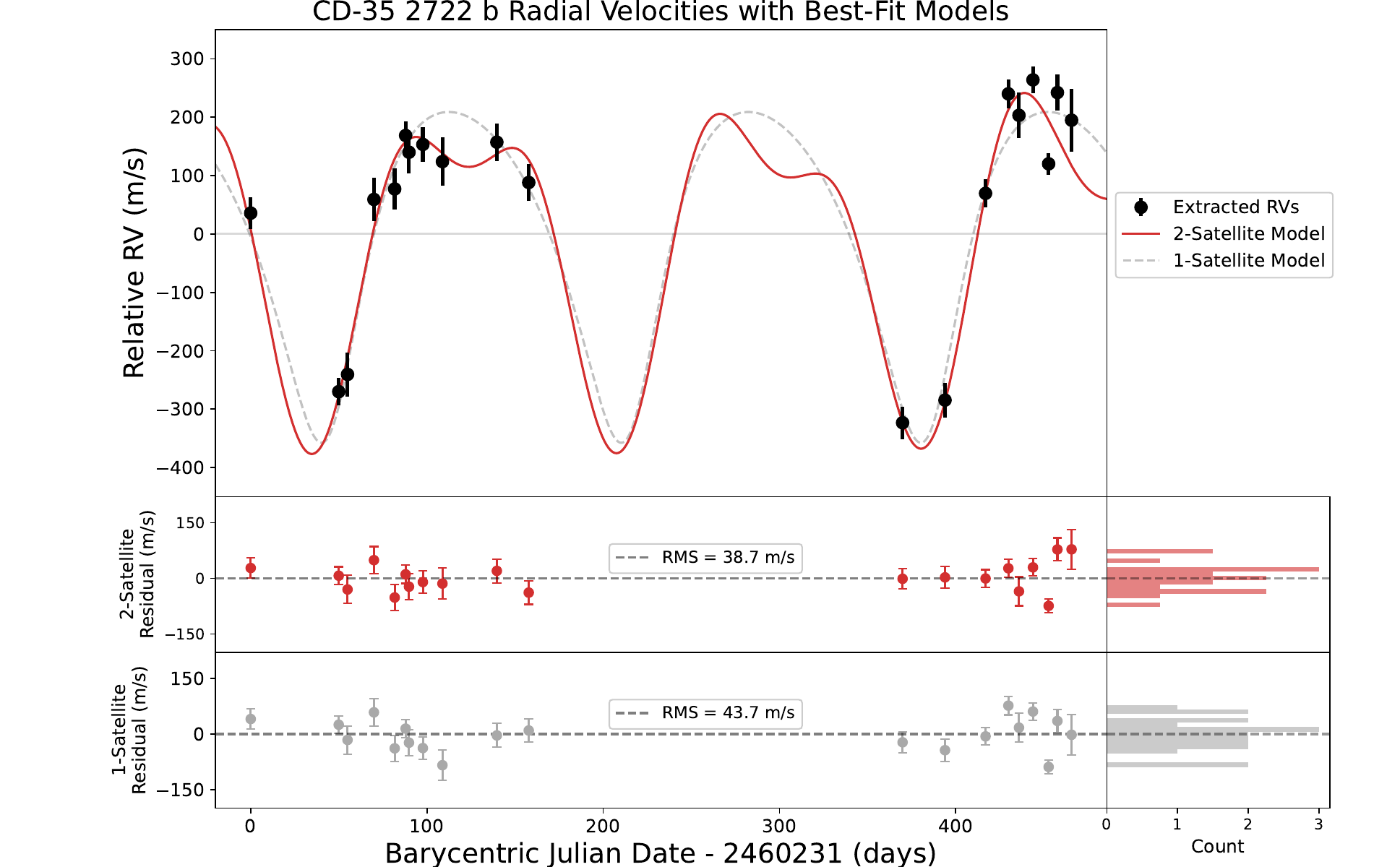}
    \caption{
        Extracted relative RVs for CD\mbox{-}35 2722 B with best-fitting one-satellite and two-satellite models. All BJD, RV, and RV error values are available in Appendix A, with the full corner plot for the favored model in Appendix B. 
    }
    \label{fig:orbit}
\end{figure}

Beyond observing a strong periodicity in our data, we use the \texttt{EMPEROR} code\cite{pena2025} for Keplerian fitting of the RV data, presented in Fig.~\ref{fig:orbit}. We are able to fit our data with different models invoking different numbers of satellites. \texttt{EMPEROR} iterates through these models until the current iteration fails to meet a fit-quality criterion, at which point the previous is accepted as the best-fitting model.

%The best-fitting model has 2 satellites, both on nearly circular orbits and nearly in mean-motion resonance (see Fig.~\ref{fig:orbit}). The periods of LONGPERIOD and SHORTPERIOD for the larger and smaller satellites, respectively, are very near a 2:1 ratio. This same effect is seen in the Galilean moons of Jupiter, with Io, Europa, and Ganymede in a 4:2:1 resonance, and could be common for satellite systems around large bodies. Mean-motion resonances and circularized orbits are mutually supportive \cite{ketchum2011}, which are also derived in our fit, demonstrating self-consistency with this model.  

The best-fitting models have 2 satellites. These models always involve a large satellite with a period of $\sim$169 days (in agreement with the periodogram), with an m$\sin{i}$ of 0.74 \MJup. Its orbit is almost circular, with an eccentricity $<$0.01. We subtract this model from the data, and show the GLS periodogram for the residuals in the third panel of Fig.~\ref{fig:periodogram}.

In the aforementioned residual periodogram, one can see that there is a strong signal for a secondary periodicity. However, there is not a clear period for this second satellite. There are strong peaks at approximately 14, 70, 88, and 115 days, with the 88 day period showing the strongest peak just above the 1\% FAP threshold. This periodogram is evidence that there is a secondary signal in the data, but does not give sufficient evidence for the period of that signal. This is likely due to the incomplete sampling of the entire period, coupled with the relatively few RV points available.

We find that the 88-day period model is the most favored model for the data, but it is not highly favored over the 115-day model, which is the second best-fitting model. With a difference in Bayesian evidence ($\Delta$logZ) of 2.6, the 88-day model is $\sim$14x more likely than the 115-day model. Thus we present it here as the most favored model, but more data will be necessary to express with certainty the period of the secondary signal. For more details, see the \texttt{EMPEROR} section in Methods.

Overall, in this preferred 2-satellite model, the larger satellite has an M$\sin{i}$ of 0.74 \MJup\ and the smaller has an M$\sin{i}$ of 0.277 \MJup, giving minimum mass ratios of 2\% and 0.7\% relative to their 37 \MJup\ host, respectively \cite{gratton2024}. This is quite large in comparison to moons in the Solar System. The Earth-Moon system has the highest ratio at 1.2\%, with all moons orbiting our giant planets being significantly smaller by mass ratio. From population synthesis of satellites, massive satellites are more common in the gravitational instability scenario \cite{inderbitzi2020}, which is one of the favored pathways invoked to explain the formation of giant companions at large separation, such as CD\mbox{-}35 B. Planets in the Solar System are expected to have formed with core-accretion mechanisms, which could explain the significantly lower moon mass ratios \cite{helled2014}. This most-favored secondary period is also very near a 2:1 mean-motion resonance (MMR) with the stronger signal. In the Solar System, Io, Europa, and Ganymede are in a 4:2:1 MMR, so it is possible for such resonances to be common on the smaller scale of satellite systems.

Unfortunately, we cannot make any assumption about the inclination of the orbit of the satellites and therefore, estimate their true masses. In the solar system, while all the planets orbit within the ecliptic, their moons do not. Rather, regular satellites (which includes all of the most massive moons in the solar system except Neptune's Triton) orbit in the plane of their hosts' equator, and the obliquity of CD\mbox{-}35 B is entirely unconstrained. 

We also explored models with a single satellite, but they delivered fits with significantly lower evidence. We recover a satellite almost identical to the larger, longer-period satellite from the previous fit. However, this model has the satellite on a significantly eccentric orbit (0.29). However, it has been shown that systems hosting bodies on circular orbits in 2:1 MMR are mathematically degenerate with a single body on a more eccentric orbit \cite{anglada-escude2010}. This effect has been explored in simulations, where a derived eccentricity of $\approx$0.3 is common when fitting 2:1 MMR systems \cite{wittenmyer2019}. %This effect is limited to enhancing eccentricity up to about 0.5, beyond which it is much more likely that the eccentric model is correct, but our result is well below this limit \cite{wittenmyer2019_2}. 

This eccentric single-satellite model would be a contextually interesting result for the system, though, as a high satellite eccentricity would align with the host's very high eccentricity of $>$0.9 \cite{bowler2020}. As the host-satellite pair follows its own highly eccentric orbit, the variation in gravitational force from the host star could cause an increase in the eccentricity of the satellite. For exoplanets in S-type orbits in stellar binary systems, high eccentricities are common, and a satellite in the CD\mbox{-}35 2722 system would experience similar dynamics \cite{su2021}.

While both models can fit the data reasonably well, the 2-satellite model passes the selection criterion and is significantly favored with a $\Delta$logZ of 6.9. All best-fit parameters for both models are available in Table~\ref{tab:fits}.

%The simplest best-fitting model is a a single, large moon on a 0.295 eccentricity orbit. orbits, however, the best fit becomes a 2-moon model, involving a large moon with nearly the same period and mass as the 1-moon model, but with an additional smaller moon on a shorter orbit. The full lists of best-fit parameters for both models, including statistics (how do I say 'things that make us like a fit more better like chi², evidence, etc'?), can be seen in Table~\ref{tab:fits}.

\begin{table}[h!]
\centering
\begin{tabular}{lcc}
\toprule
\textbf{Parameter} & \textbf{2-satellite Fit} & \textbf{1-satellite Fit} \\
\midrule
Period 1 (days)              & $169.45^{+1.1}_{-1.06}$ & $170.05^{+0.13}_{-0.97}$ \\
Amplitude 1 (m/s)            & $246.45^{+7.03}_{-5.02}$ & $283.27^{+1.48}_{-13.53}$ \\
Phase 1 (rad)                & $4.04^{+1.97}_{-0.1}$  & $4.58^{+0.07}_{-0.08}$ \\
Eccentricity 1           & $0.005^{+0.008}_{-0.005}$   & $0.29^{+0.042}_{-0.004}$ \\
Longitude of Periastron 1 (rad) & $4.07^{+0.02}_{-2.05}$ & $3.55^{+0.07}_{-0.07}$ \\
Semi-Major Axis 1 (au)          & $0.199^{+0.005}_{-0.004}$ & $0.196^{+0.003}_{-0.001}$   \\
M$\sin{i}$ (\MJup)          & $0.743^{+0.005}_{-0.039}$ & $0.778^{+0.034}_{-0.014}$ \\
\midrule
Period 2 (days)       &  $87.46^{+0.0}_{-0.63}$     & --- \\
Amplitude 2 (m/s)     &  $113.92^{+14.07}_{-14.1}$    & --- \\
Phase 2 (rad)         & $1.4^{+0.45}_{-1.39}$    &  --- \\
Eccentricity 2        & $0.01^{+0.01}_{-0.01}$     &  --- \\
Longitude 2 (rad)     & $5.59^{+0.7}_{-5.59}$     &  --- \\
Semi-Major Axis 2 (au)    &  $0.129^{+0.003}_{-0.003}$    & --- \\
M$\sin{i}$ (\MJup)      & $0.277^{+0.035}_{-0.042}$     & --- \\
\midrule
Offset (m/s)               &  $-77.73^{+7.33}_{-1.15}$ & $-59.86^{+3.26}_{-5.27}$ \\
Jitter (m/s)               &  $16.39^{+5.34}_{-5.90}$  & $24.16^{+4.47}_{-4.4}$ \\
\midrule
logZ                       &  $-122.654 \pm 0.952$ & $-129.295 \pm 0.920$ \\
\bottomrule
\caption{Best-fit orbital parameters for one- and two-satellite models. Period, amplitude, phase, eccentricity, and longitude of periastron are the components of a Keplerian equation. Offset is not a physically meaningful parameter, it is a constant offset subtracted from the model to align the model and RV points. Jitter is a parameter that models instrumental RV errors, which is then added in quadrature with the measured RV errors.}

\end{tabular}
\label{tab:fits}
\end{table}

%\section{Discussion}

To verify the plausibility of this detection, we calculate the Roche Limit, within which a satellite would be torn apart by tidal forces, and the Hill Radius, outside of which the host star's gravity would dominate, to determine if the moons would be stable. 

The Roche limit depends on the radius of the host, and the densities of both host and satellite. Using isochrones with CD\mbox{-}35 B's age and mass, we find a radius of 1.2 \RJup, and thus 29 g/cm$^3$ for the mean density \cite{baraffe2015, gratton2024}. For the presented satellites, we conservatively estimate that they both have the same density as Saturn (0.687 g/cm$^3$). It is impossible to say how accurate this approximation is, as we have no information regarding the satellite radii, but this should be a realistic underestimate of the true densities. If the satellites are beyond this upper limit on the Roche limit at this underestimated density, they must also clear this criterion at their true densities. We calculate a maximum Roche limit of 8.4 \RCD, well within the proposed 356 \RCD\ (0.2 au) semi-major axis for the large moon, and the 231 \RCD\ (0.13 au) semimajor axis for the smaller. These orbits are allowed and the satellites could not be disrupted by tidal forces.

The Hill Radius for CD\mbox{-}35 B varies significantly during its orbit, as it scales with distance to the host star. The stability limit for satellites is not the Hill Radius itself, but can be expressed as a fraction of the Hill Radius, depending on the inclination of the orbit relative to the star and the eccentricities of the planet's and satellite's orbits \cite{domingos2006}. Beyond this limit, the perturbation of the host star will eventually cause satellites to be ejected or spiral into their host. A coplanar, prograde orbit is the most limiting case, with a stability limit in units of the Hill Radius given by equation 5 in Domingos et al. (2006)\cite{domingos2006}:
\[
a_E \approx 0.49\left(1 - 1.0305\,e_{\text{planet}} - 0.2738\,e_{\text{satellite}}\right)
\]
Again, our model orbits are fully compatible with the stability limit, which is 1.07 au (both satellites have such low eccentricities, they yield the same result). This is much larger than any derived orbits. For such a large companion orbiting a relatively small star at a long period, this large stability limit is the expected result. 

Another concern for the reality of these detections is the rotation of the host. While CD\mbox{-}35 B does not have the same kind of spots or chromospheric activity of stars, it does likely have clouds which could experience circulation, driving brightness variations over time. This effect could drive RV variation on the rotation period of the BD, or on the timescale of cloud restructuring in the atmosphere, although there is some doubt that there is substantial cloud coverage in this object \cite{wang2025}.

Importantly, the target is expected to rotate much faster than late-type stars tend to, due to its significantly smaller radius. Unfortunately we cannot directly measure the rotation period from our data and no prior measurements exist. The v$\sin{i}$ has been measured to be 9.58 km/s, giving a maximum rotation period of 0.65 days\cite{wang2025}. This is significantly shorter than the long-period signal derived in this work. It also demonstrates that there is likely not a large rotational signal embedded in the data, as we do not see any strong RV changes on that timescale. Moreover, we compare CD\mbox{-}35 B to other BDs, to demonstrate that these results are reasonable, and a rotation period of $>$100 days would be an extreme outlier. 

In the AB Doradus association, of which the CD\mbox{-}35 2722 system is a member, 9 BDs were measured to have rotation periods of at most 18 hours \cite{vos2022}. The BDs in the 1-Myr-old Taurus star-forming region show a positive trend in a period-mass relation, with the slowest rotators being $\sim$5 days for BDs $\sim$3x more massive than CD\mbox{-}35 B \cite{scholz2018}. For another target of similar age, CFHT-PL8 in the Pleiades, age $\sim$100 Myr, the period has been measured to be 0.4 days \cite{terndrup1999}. 

These results strongly disfavour rotation periods for CD\mbox{-}35 B longer than a few days, with a period of $\lesssim$0.65 days being reasonable. A slower rotation would imply a correspondingly low equatorial velocity, making it highly improbable that rotation could produce the observed strong signal, which has an amplitude of $\sim$500 m/s. Although long-timescale variability driven by magnetic activity cycles or large-scale atmospheric circulation cannot be ruled out in principle, such mechanisms are unlikely to generate RV variations of this magnitude.

Rotationally induced variability remains a potential concern. Objects with spot filling factors of only a few per cent can produce radial-velocity modulations of several hundred meters per second \cite{vanderburg2018}. However, on short timescales we do not detect excess jitter in our data at levels comparable to the inferred signals. The observations are irregularly sampled on timescales of both hours and days, making it unlikely that rotational modulation would consistently be observed at similar phases; strong rotational signatures should therefore be apparent if present, yet no such scatter is detected. We therefore consider large-amplitude rotational effects improbable for the dominant signal. By contrast, the secondary, shorter-period signal could more plausibly arise from lower-amplitude rotational variability. An $\sim$88-day period would still be unusually long for rotational modulation, but we nonetheless treat this second signal with greater caution than the first.

%\section{Conclusion}

Ultimately, this is the most compelling detection of a satellite around a substellar companion to date. At a minimum, the apparent RVs of CD\mbox{-}35 B show strong periodicity. None of our tests have found other mechanisms that could induce the derived RV variations, making it very likely that these are genuine satellite signals. With future follow-up observations, we hope to further confirm the detections and better characterize the orbits of our two candidates. 

The discovery and characterization of satellites will open many new lines of investigation in the realms of exoplanetary science. Planet formation theory already makes some predictions regarding satellite formation and statistics which can be refined with comparison to a real sample. Dynamicists will have access to a new dimension of system hierarchical structure. Although those discovered in this work are likely too massive to be viable hosts for it, satellites are also fascinating targets in the search for life, as they can receive tidal heating from their host planet, potentially allowing them to be habitable beyond classical stellar habitable zones \cite{heller2014}.

The nature of this system prompts an important question: how do we define an exomoon? Their orbits around a substellar host, mass ratios of $<$1\%, and position as the third component in a hierarchical stellar system all support a moon-like classification. Conversely, both the satellites and their primary are unusually massive, placing them in regimes that blur the distinctions between moons, planets, and stars. 

The current IAU definition of planet states, "Objects with true masses below the limiting mass for thermonuclear fusion of deuterium (currently calculated to be 13 Jupiter masses for objects of solar metallicity) that orbit stars, brown dwarfs or stellar remnants and that have a mass ratio with the central object below the L4 / L5 instability $(M/M_{\mathrm{central}} < 2/(25+\sqrt{621}) \approx 1/25)$) are 'planets', no matter how they formed"\cite{lecavelier2022}. This definition makes no reference to system hierarchy and explicitly states that planetary-mass objects orbiting brown dwarfs are planets. However, the CD-35 2722 system brings into stronger relief the difference in the future evolution between B's satellite and a hypothetical planet orbiting A. When this system reaches the same age as the current Solar System, the BD is expected to be approximately two orders of magnitude, while the central star continues shining, as bright as it is now\cite{marley2021, dieterich2021}. This difference between a stellar host and a substellar host has deep ramifications for the satellite itself, and it may be worth revisiting this aspect of the definition.

At present, there is no accepted definition of an exomoon, and a single criterion may prove inadequate, although this classification may also not be apt for this satellite. Perhaps we are approaching the limit of language invented to describe the Solar System, whose architecture is entirely different from that of CD-35 2722. In which case, it may be best to define a new category for objects like the satellite of CD-35 2722 B, including some objects orbiting BDs. %A classification scheme that distinguishes, for example, “gaseous moons” from the rocky satellites of the Solar System may be more appropriate. 
With forthcoming observational facilities expected to uncover many similar systems, establishing a clear taxonomy will become increasingly important. 

\newpage
\section{Methods}

\subsection{Observations and Data Reduction}

The data presented are part of a survey conducted with VLT/\CRIRES, a high-resolution cross-dispersed infrared echelle spectrograph, recently upgraded to greatly increase wavelength coverage, incorporate new calibration elements, and add a variety of other features \cite{dorn2023}. We have obtained 21 epochs of observations of CD\mbox{-}35 B. Of these, one is excluded from the analysis due to a very low signal-to-noise ratio (S/N) in the spectral continuum of $\sim$5. The spectra have a resolution of $\sim$100,000 and span a wavelength range between 1469 and 1780 nm. The observations took place over the span of fifteen months, from October 2023 through January 2025. 
In order to achieve the S/N necessary ($\sim$25) for this analysis, we required 10-minute exposures. They were observed in an AB nodding pattern in order to subtract the sky emission contamination that is especially strong with such long exposure times, as well as the instrumental background.

To minimize the potential for stellar contamination, the slit was aligned to be perpendicular to CD\mbox{-}35 B's orbital vector, maximizing the distance between the host star and any given position on the slit. Adaptive optics is also employed to sharpen the target point-spread function and further reduce contamination. Finally, because CD\mbox{-}35 B is on such a wide-separation orbit, there should be very little reflected light from the host star compared to its intrinsic brightness. Any potential contamination would also be reduced by the nodding subtraction, although the increased distance from the host star to the other nodding position limits the potential correction. Combining these factors, the stellar contamination is likely negligible and does not need to be modeled together with the companion's signal.

To quantitatively demonstrate the lack of contamination in our data, we fit model point-spread functions (PSFs) to the slit viewer (SV) camera images associated with our data. For our observation strategy, we guide the telescope on the host star, while offsetting the spectrograph slit to the known position of the companion. The SV images show the PSF of the host star, as reflected off of a panel surrounding the instrument's two slits. To model the host star PSF, we fit a 2D Moffat function. This function is the best approximation of the shape of the AO-corrected PSF.

\begin{figure}[t]
    \centering
    \includegraphics[width=\linewidth]{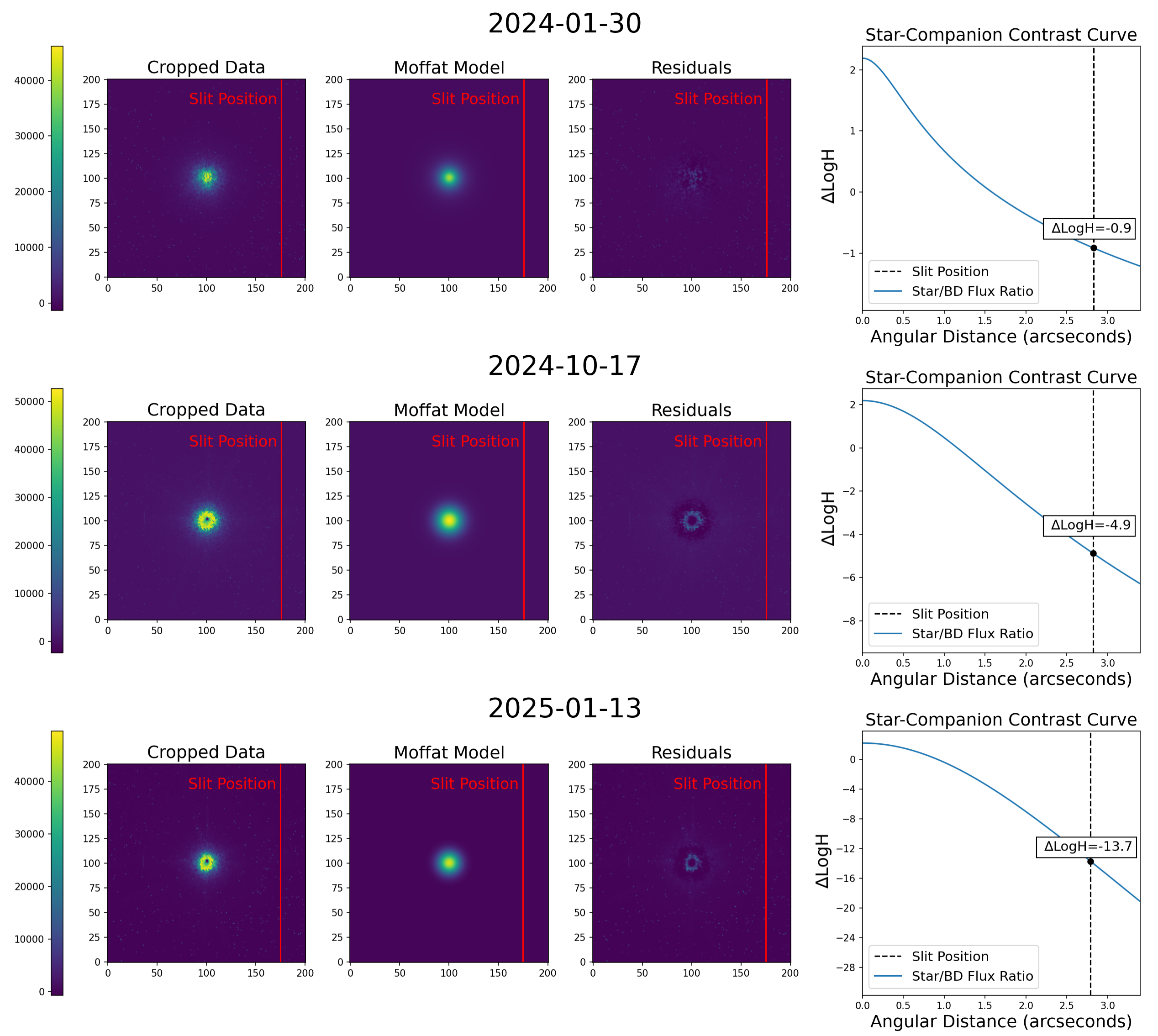}
    \caption{
        Selected SV images associated with our spectra, fit with Moffat model functions. The contrast curve in the rightmost panels shows the log-scale H-band flux ratio ($\Delta$LogH) between the host star and CD\mbox{-}35 B as a function of angular distance from the star, specifically marking the position of the slit. This is based on the relative diminishment of the model and the contrast of the star and companion given in the discovery paper \cite{wahhaj2011}. The first row is the night with the worst observing conditions, and shows a maximum contamination of $\sim$15\%.  The other two selected epochs are under better weather conditions. The dark pixels in the center of the real PSF are due to saturation of the detector. These points are masked when fit, but not having central pixels does complicate the fit, as can be seen in the residual ring for the two good weather observations. This effect should not significantly affect the results at separations as distant as the slit.
    }
    \label{fig:svfits}
\end{figure}

Fig.~\ref{fig:svfits} shows the PSF fits to the SV camera images along with a 1D radial projection to measure the contamination as a function of angular distance. The night with the worst conditions was January 30, 2024 (seeing up to 1.73 arcsec). In that worst case scenario, the potential flux ratio between the companion and the contamination is $\sim$13\%, excluding that which is subtracted in the nodding. The effect of this can potentially be seen in Fig.~\ref{fig:orbit}. This epoch is the third to last point in the first year of observations, which has slightly larger error bars and is lower than the neighboring points. Even assuming this reduction in measured RV is due to the contamination, this would demonstrate that its influence is, at worst, some tens of meters per second, far lower than the measured amplitude of either signal. This PSF is also significantly worse than the average, with all other points having an apparent contamination orders of magnitude lower.

To reduce the data, we used the official ESO pipeline, using the cr2res recipes (version 1.6.10)$^[$\footnote{https://www.eso.org/sci/software/pipelines/cr2res/cr2res-pipe-recipes.html}$^]$. This reduction was performed using the default settings of the pipeline. 

When reducing nodding observations, it is common practice to not only subtract the nodding frames to remove sky contamination, but also to combine the extracted spectra into a single combined spectrum in order to improve S/N. In this work, however, we choose to use the individual nodding frames as separate observations through the RV extraction phase, after which they are binned into a single RV measurement per night of observation. This is motivated by the need for a wavelength solution that maintains the maximum possible accuracy. \CRIRES is not a stabilized instrument, so it is not guaranteed that the wavelength solutions between the A and B nodding frames are the same, and minute-to-minute drifts are known to exist within the instrument. Therefore, we choose to eschew the S/N gain of spectral combination for the ability to recalibrate the wavelength solutions separately for the two nodding positions. 

\begin{figure}[t]
    \centering
    \includegraphics[width=\linewidth]{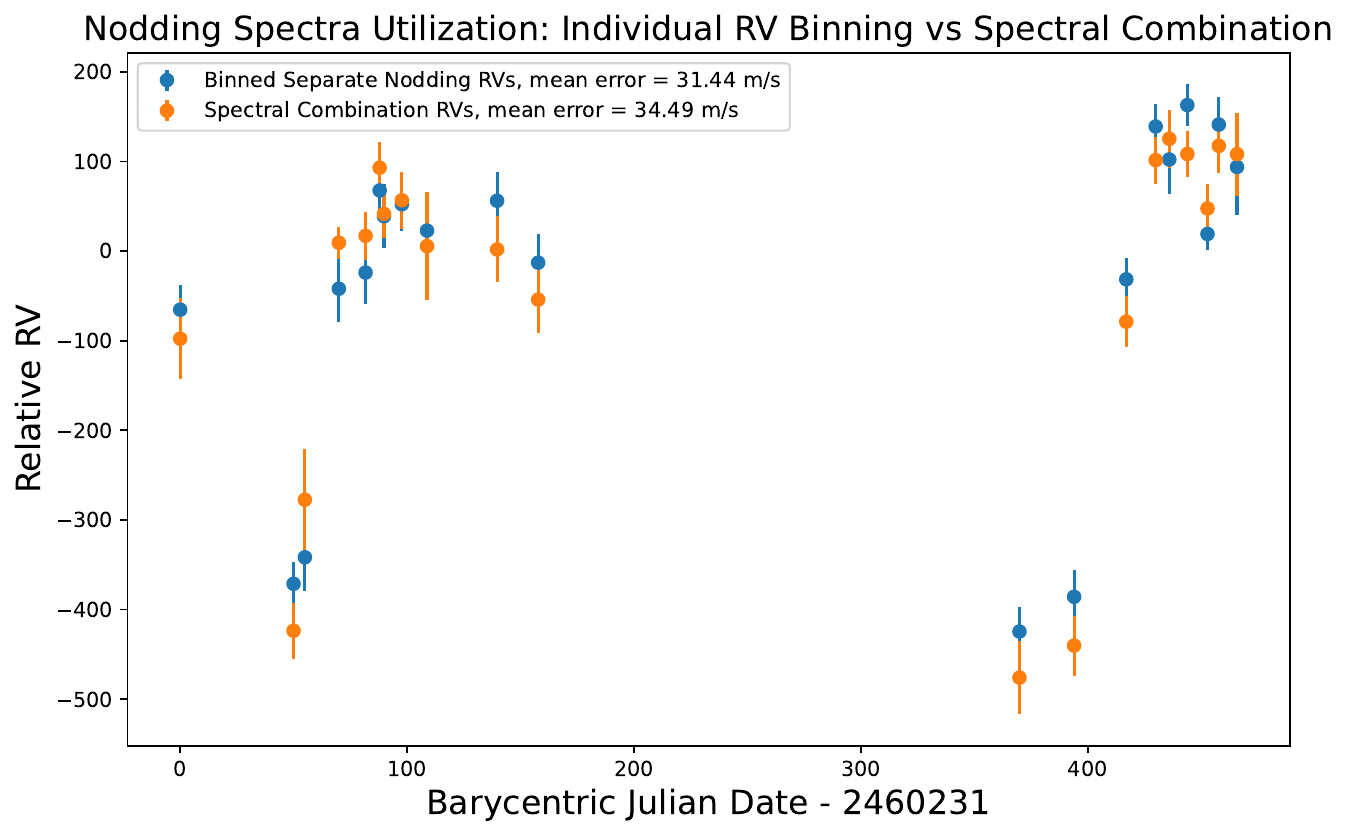}
    \caption{
        Comparison of final RV calculation methods. The blue points are the favored method, in which we calculate separate RVs for both nodding positions, then bin the results. The orange points are based on combining the two nodded spectra before calculating the RV. The precision improvement with the former method is $\sim$10\%, as seen in the mean error in the legend. 
    }
    \label{fig:nodding}
\end{figure}

In short, by using the frames separately, we lose RV precision through S/N loss, but gain higher-precision wavelength calibrations for each frame. Then, assuming the true RV of the target is constant over the short time between the two nodding positions, we can bin these two measurements with a weighted mean and associated error propagation to produce a single, higher-precision RV measurement for each night. The precision gain of this method is approximately 10\%, as can be seen in Fig.~\ref{fig:nodding}.

\subsection{Radial Velocity Extraction: \texttt{viper}}

%To extract the RV's, we used the template-matching code \texttt{viper} \cite{zechmeister2021}. The code combines all observed spectra together to create a template that can be used to calculate the relative RV between different observations. This is done iteratively, with the first template being made by subtracting the barycentric velocity from each frame. Then, using this first template, one can get an estimate of the true RV variation, apply that correction, and produce a more precise template. The results in this work are based on two template iterations, as subsequent iterations showed little to no change in the extracted RVs. 

%\texttt{viper} also performs the wavelength calibration of the data. It aligns the telluric lines in the data with those from theoretical models, which are stable to about 10 m/s \cite{figueira2010}. Once the tellurics are aligned, it corrects the spectrum to remove the contamination before RV calculation and template creation. 

To extract the RVs, we use the RV pipeline \texttt{viper}\cite{zechmeister2021, kohler2025} which employs a forward-modeling approach. The code offers the option to co-add all spectra to create a telluric-free, high S/N template of the target. This template serves as a reference for the calculation of the RV values. The creation of the template is an iterative process. In the first step, the telluric-corrected spectra are co-added after shifting them into the same barycentric rest frame. In a next step, this template can be used as an input for a second round of template creation, allowing the code to correct for RV shifts between the different observations, improving the next template. The results in this work are based on two template iterations, as subsequent iterations showed little to no change in the extracted RVs.

Since the parameters vary strongly with wavelength, the RVs are calculated separately for each spectral order. The final RV is the weighted mean of the individual RV values of all spectral orders. The corresponding final uncertainty is the standard deviation of the individual order RVs, weighted by the individual errors and divided by the square root of the number of orders included. The errors of the individual orders are provided from the covariance matrix of the least-square fitting. For more detail regarding the RV calculation, see Kohler et al. (2025)\cite{kohler2025}.

In \texttt{viper}, all parameters are optimized simultaneously, including the wavelength solution. In our case, it uses the telluric lines as a reference frame, as these lines are expected to be stable to about 10 m/s\cite{figueira2010}. However, not all orders contain sufficient telluric lines for a high-quality recalibration. These orders are excluded from the calculation, as this failed calibration produces highly erratic results. %However, as this is only the standard deviation, it is possible that even greater RV shifts may occur between different epochs.

%In another data set, which is similar to the one used here but will be analyzed in a future work, \texttt{viper} calculates an overall RV variation of $\sim$160 m/s. If we assume that there is no physical effect causing RV variation in that other target, and that the true RV precision of \texttt{viper} for this data is 160 m/s. This is still well below the observed RV variation. Thus, \texttt{viper} is entirely capable of detecting the 500 m/s signal of the larger satellite. However, the secondary 220 m/s signal is not significantly large relative to this conservative error estimate, and it is possible that it is an artifact of some unknown systematic. Still, we have no reason to distrust the calculated errors, which are well below the signal amplitudes.

\subsection{Keplerian Orbit Fitting: \texttt{EMPEROR}}

After obtaining the RVs from \texttt{viper}, they are given as input to the Keplerian orbit-fitting code \texttt{EMPEROR} \cite{pena2025}. This code employs multi-tempered MCMC fitting to quickly probe the large parameter space of possible orbits. It iterates through Keplerians of different degrees (i.e., one orbiting body, two orbiting bodies, etc.) until the Nth Keplerian is not a significantly better fit than the last.
For a deeper explanation of the underlying mathematics and algorithm of \texttt{EMPEROR}, see Peña et al. (2025)\cite{pena2025}.

When using \texttt{EMPEROR}, we use a different parameterization of the orbital elements than the one presented in the results. Instead of directly probing the eccentricity ($e$) and longitude of periastron ($\omega$), as reported, we fit $e\sin{\omega}$ and $e\cos{\omega}$. This generally favors lower-eccentricity orbits and prevents the model from overfitting the data with high-eccentricity, short-period orbits in higher order Keplerians. High-eccentricity orbits tend to be capable of fitting a Keplerian signal to any given dataset, even where none exists.  We convert these parameters back to eccentricity and longitude of periastron when reporting them, as those quantities are more easily conceptualized independently.

For the initial run, we employ wide priors to avoid biasing the fit. From that run, we identify strong degenerate peaks in the posteriors, as the code cannot define which signal should be "satellite 1" or "satellite 2", just that both signals exist. We then narrow the period prior to "satellite 1" to only include the strong, long-period peak, limiting "satellite 2" to shorter-period signals.

\begin{figure}[t]
    \centering
    \includegraphics[width=\linewidth]{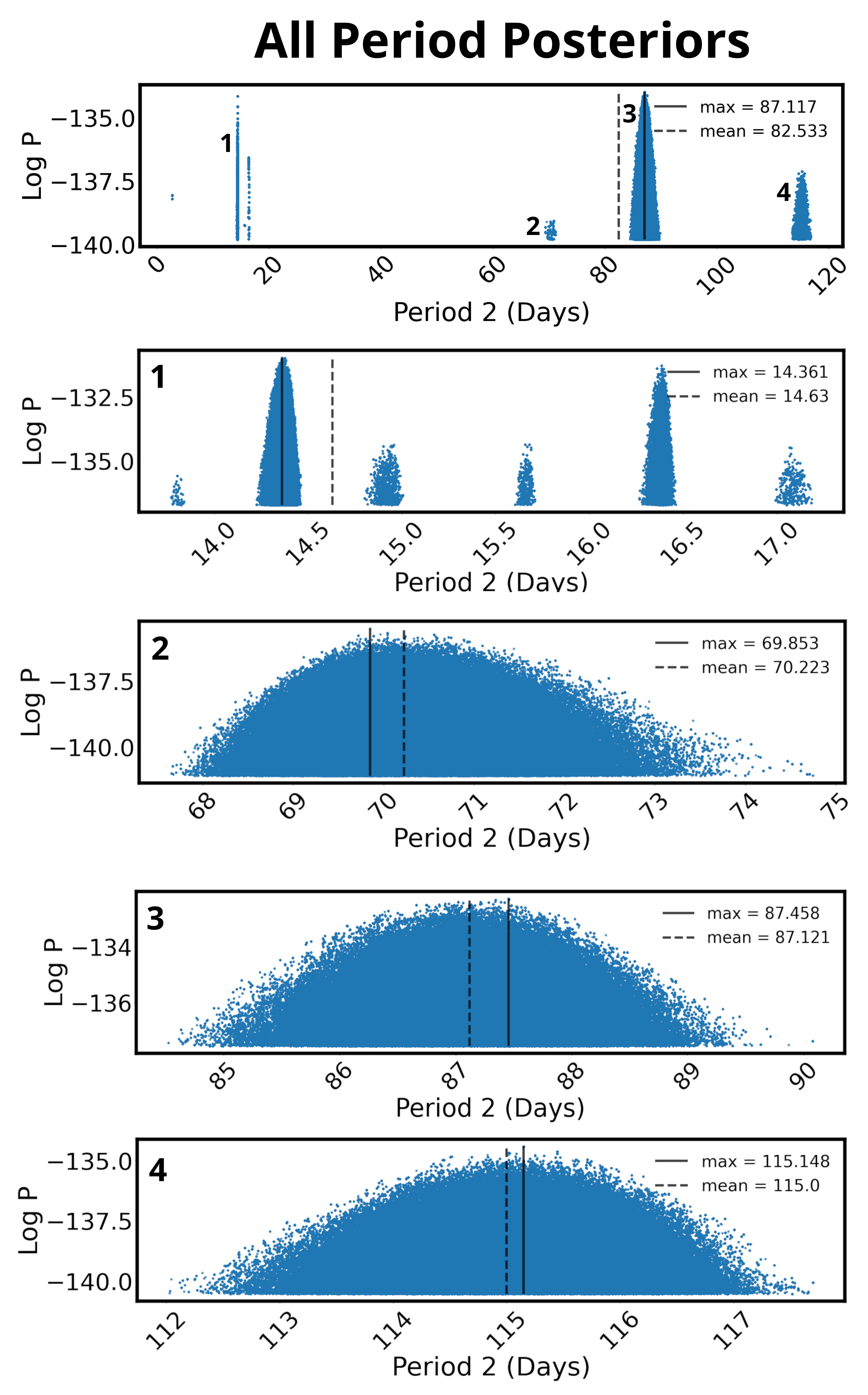}
    \caption{
        Period posteriors for the main and windowed 2-satellite model fits. The first panel shows all the periods recovered for the second signals. The other panels are windowed fits for each of the peaks seen in the first panel, and are labeled 1-4 accordingly.
    }
    \label{fig:multiposts}
\end{figure}

When we separate these signals into greater than or less than 150 days, we always recover the strong $\sim$170 day period in the long regime. However, in the short-period regime, there are degeneracies in the posterior, showing multiple potential secondary signals. These posteriors are in the first panel of Fig.~\ref{fig:multiposts}. In order to compare which model fits best, we break these posteriors into multiple windows covering each period separately in order to get a Bayesian evidence estimation for each solution. There are 4 possible solutions at periods of 14 days, 70 days, 88 days, and 115 days. These periods are all aliases of each other with our current sampling, due to the two sets of observations being almost exactly a year apart. Observing this target again, avoiding this $\sim$1 year mean time difference will help break this degeneracy.

The 14-day period model is somewhat of a special case, as the window encompassing it showed, in itself, a highly degenerate period posterior with several period solutions spanning 13 to 17 days. This leads us to believe that periods this short are simply too difficult to constrain with our current dataset. Below this $\sim$20 day period limit, there are many solutions which could potentially fit, but these are likely artifacts of our low-cadence sampling. Periods that are very short compared to the sampling of the data are often capable of resolving the residuals of an imperfect fit, which, in this case, is the 1-satellite circular orbit fit. Thus, we reject the models in this period window. 

For the other three windows, we must rely on the evidence estimations to determine the most favored model, as the posteriors are all singular and approximately gaussian in these windows. Based on the derived evidences, the 88-day model is about 14x more likely than the 115-day model, and 18x more likely than the 70-day model. While this difference is significant, it is not sufficiently strong to claim with certainty a known period for the potential secondary signal. %We are confident that a second signal is present, and have a favored model for the period of that signal, but the former claim is much stronger than the latter. 

\begin{figure}[t]
    \centering
    \includegraphics[width=\linewidth]{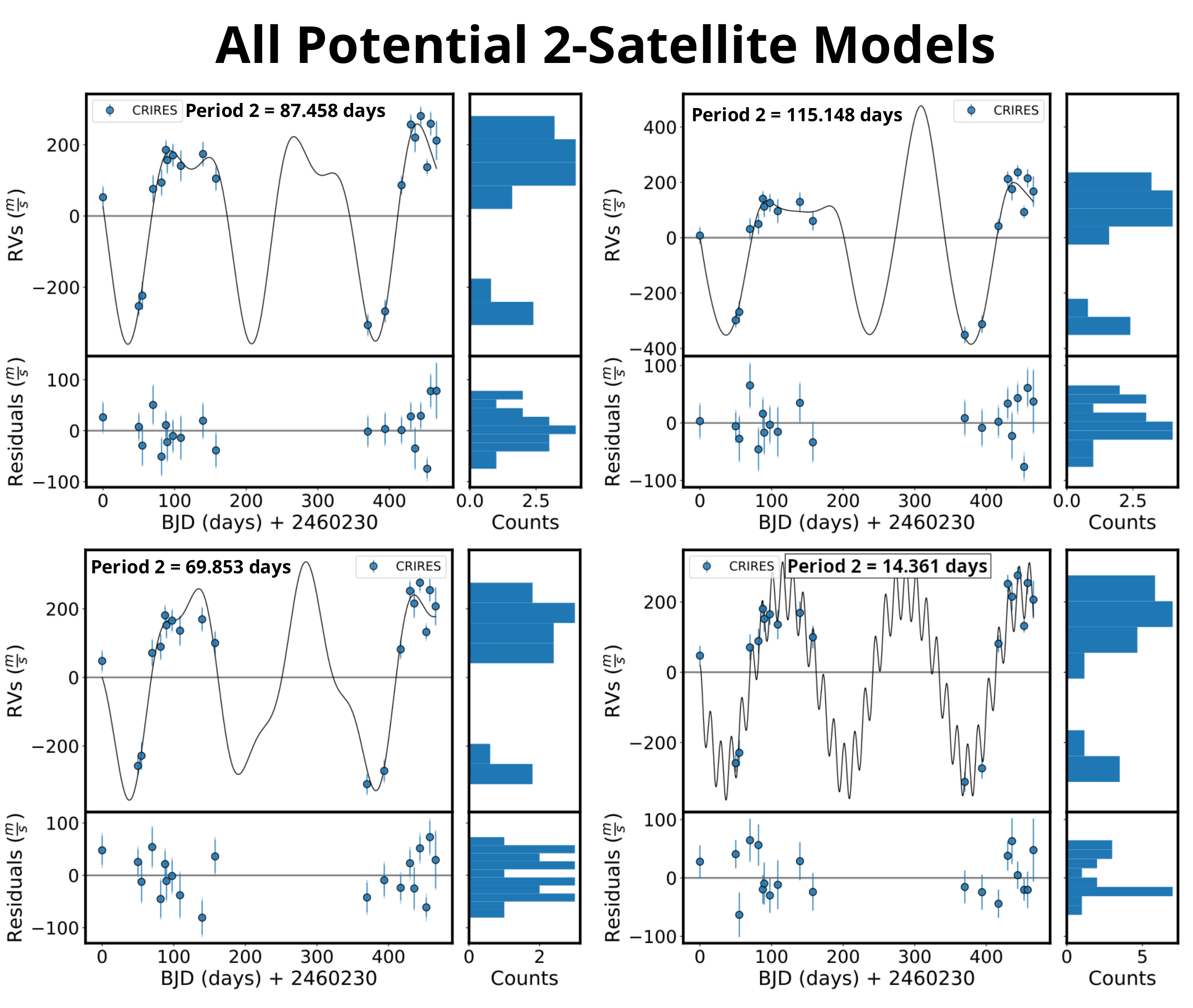}
    \caption{
        All high-evidence 2-satellite models. One can see the reason for distrusting the 14-day model, as these high-frequency oscillations would appear more likely to be an artifact of the sampling, rather than a true signal we could recover from this data.
    }
    \label{fig:multimodel}
\end{figure}

The full posteriors for all windowed fits are shown in Fig.~\ref{fig:multiposts}. The best-fit models and their residuals are shown in Fig.~\ref{fig:multimodel}. Also, to guarantee that the partially contaminated epoch discussed above was not driving a spurious detection of the second signal, we fit the data with EMPEROR excluding that epoch. The results were not significantly changed by the exclusion of that data point, and so it is left in for completeness.

%We run \texttt{EMPEROR} with two sets of priors, differing only in the constraint on eccentricity. The eccentricity parameter produces the most meaningful disparity in our results, as all models we fit have agreed on other parameters (period, amplitude, etc.) Forcing circular orbits produces a significantly different result, as will be discussed in the following section. The priors full sets of priors can be seen in Table~\ref{tab:priors}

\subsection{Verifying the periodicity of the signal}

On initial inspection, the period of the larger satellite could be seen as suspicious, as it is relatively near 182.5 days, half an Earth year. If there are imperfect corrections applied to, say, the barycentric velocity or instrument profile\cite{latouf2022}, this could create a false signal with a true period of 365 days, which could be aliased to the 182.5 days by sampling. However, we have extensively investigated this possibility and have found no correlations between the derived RVs and the applied barycentric correction. We have also looked at atmospheric parameters (seeing, precipitable water vapor, etc), which could also be periodic in yearly weather patterns, and again found no correlations with our data. The derived large satellite period is also very well-constrained to  $169.45^{+1.1}_{-1.06}$ days (in the 2-satellite model), meaning 182.5 days is off by approximately 11$\sigma$. Thus, while this period could seem suspicious, it is highly improbable that it is due to any unaccounted for Earth-based effect.

\begin{figure}[t]
    \centering
    \includegraphics[width=\linewidth]{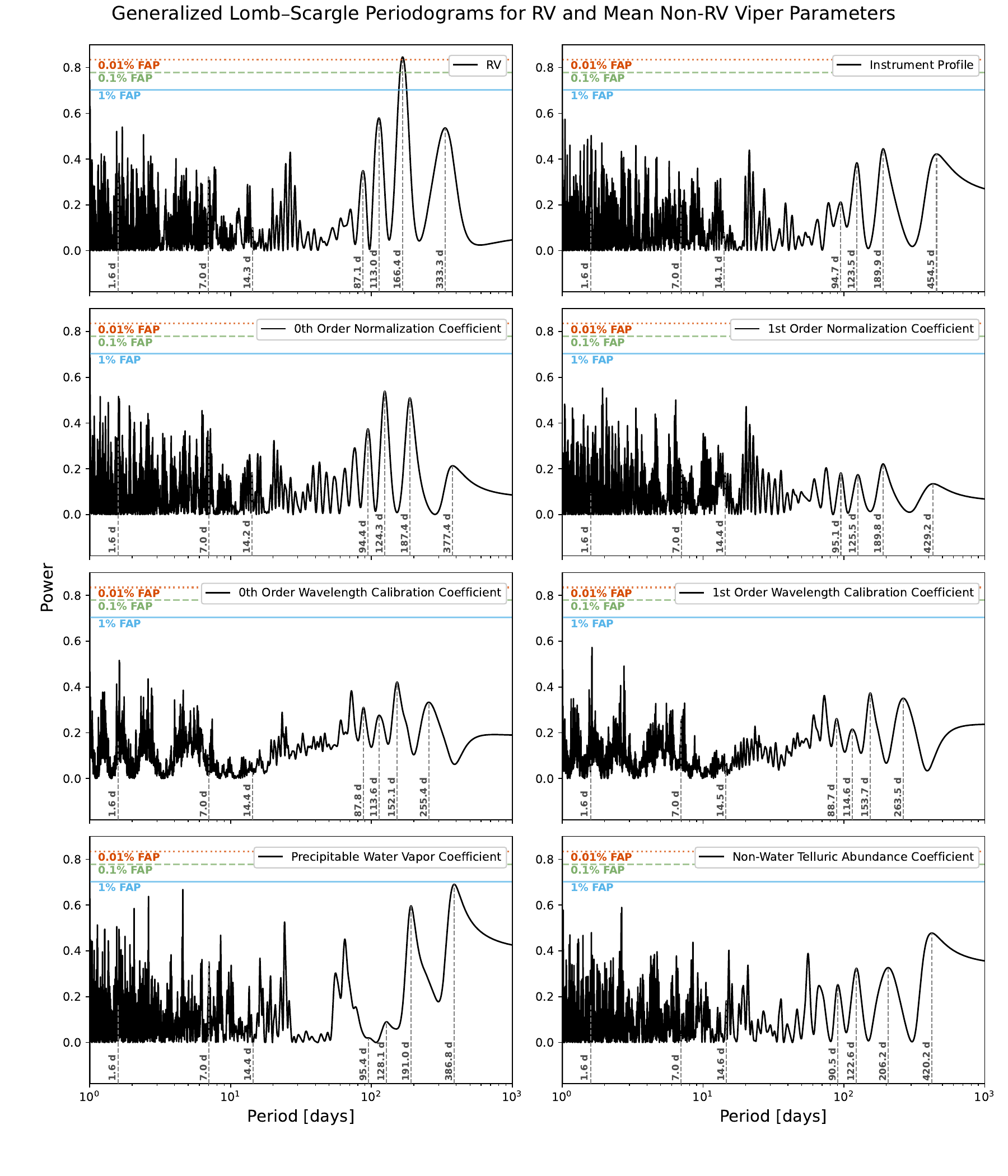}
    \caption{
        GLS periodograms for all \texttt{viper} outputs. The first panel is the same as Fig.~\ref{fig:periodogram} above, and is reiterated for comparison. The spectra have been normalized and wavelength calibrated with 2nd order polynomials, hence the paired coefficients, but the 2nd order coefficients in both cases do not significantly vary, and are thus omitted for brevity. The dashed lines in each panel are the maxima located nearest to those shown in the RV panel, to show that none repeat the same periodicities. 
    }
    \label{fig:extraperio}
\end{figure}

Further, we compared our results to GLS periodograms made using the model parameters fit by \texttt{viper} in conjunction with the RVs, to see if they shared any periodicity with the RVs themselves. No correlation can be found, with no significant periodicities present in these parameters, and even the insignificant peaks do not align with the derived periods of the satellites. These periodograms can be compared in Fig.~\ref{fig:extraperio}. Specifically, the "instrument profile" periodogram could be influenced by the rotation of the target. We assume that the primary contributor to the line profile is instrumental, hence the name, but it is theoretically possible that if there are real changes in the line profile of the object, they would manifest here. However, no periodicity is recovered for this variable, meaning it is unlikely for rotational modulation of line profiles to be a significant driver of the RV variation. Of these periodograms, the one with the strongest peaks is the one for precipitable water vapor, but those peaks are much nearer to an Earth year and its aliases, denoting seasonal weather cycles, rather than the periods of our proposed satellites.

There are four specific data points that are key to this analysis, which constitute the limited sampling of the lower end of our orbital fits. From our examination, there is nothing unique about these points beyond their low RV. They were all taken in excellent weather conditions, as were many of the points with higher RVs. The points have no outlying instrumental parameters, such as detector temperature, grating position, S/N, or telescope position. These points also come in pairs, with each point having a correlated low RV point as well. If we exclude any one of these points, we recover the exact same signal because of the neighboring point containing the same information as the one removed. It is also the case that there are actually 8 points associated with this low-RV portion of the phase, as pre-combination, the separated nodding position RV points are in agreement. While a more complete orbital sampling would be ideal, we have found no reason to doubt the validity of these four crucial points.

Lastly, we reiterate that the rotation period of CD\mbox{-}35 B is short relative to the recovered satellite periods, with a maximum period of approximately 0.65 days \cite{wang2025}. If the RV signal were driven by rotational changes in the appearance of the target, one should observe such effects over short timescales, which we do not.

\bibliography{references}

\newpage

\section*{APPENDIX}
\addcontentsline{toc}{section}{APPENDIX}
\appendix
\section{Full RV dataset}

\begin{table}[h!]
    \centering
    \begin{tabular}{c c c}
    \hline
    \textbf{BJD} & \textbf{Relative RV (m/s)} & \textbf{RV Error (m/s)} \\
    \hline
        2460230.8655 & 42.5 & 27.5 \\
        2460279.9442 & -263.2 & 24.0 \\
        2460284.9378 & -233.7 & 37.9 \\
        2460299.9224 & 65.9 & 36.7 \\
        2460311.6920 & 83.8 & 35.2 \\
        2460317.9016 & 175.5 & 24.4 \\
        2460319.8100 & 146.8 & 35.8 \\
        2460327.6805 & 160.1 & 29.7 \\
        2460338.7864 & 130.7 & 41.5 \\
        2460369.6677 & 164.0 & 32.2 \\
        2460387.7580 & 94.9 & 31.9 \\
        2460599.8891 & -316.5 & 27.9 \\
        2460623.9639 & -277.7 & 29.7 \\
        2460646.9657 & 76.4 & 23.8 \\
        2460659.9203 & 246.8 & 25.1 \\
        2460665.9276 & 210.0 & 39.0 \\
        2460673.8730 & 270.7 & 23.2 \\
        2460682.7670 & 126.9 & 18.4 \\
        2460687.7443 & 249.0 & 31.1 \\
        2460695.7760 & 201.7 & 53.9 \\
    \hline
    \end{tabular}
    \caption{
        Barycentric Julian dates, radial velocities, and their correlated errors used in this work.
    }
    \label{tab:allvals}
\end{table}

\newpage
\section{Full 2-satellite model fit corner plot}

\begin{figure}[htbp!]
    \centering
    \includegraphics[width=\linewidth]{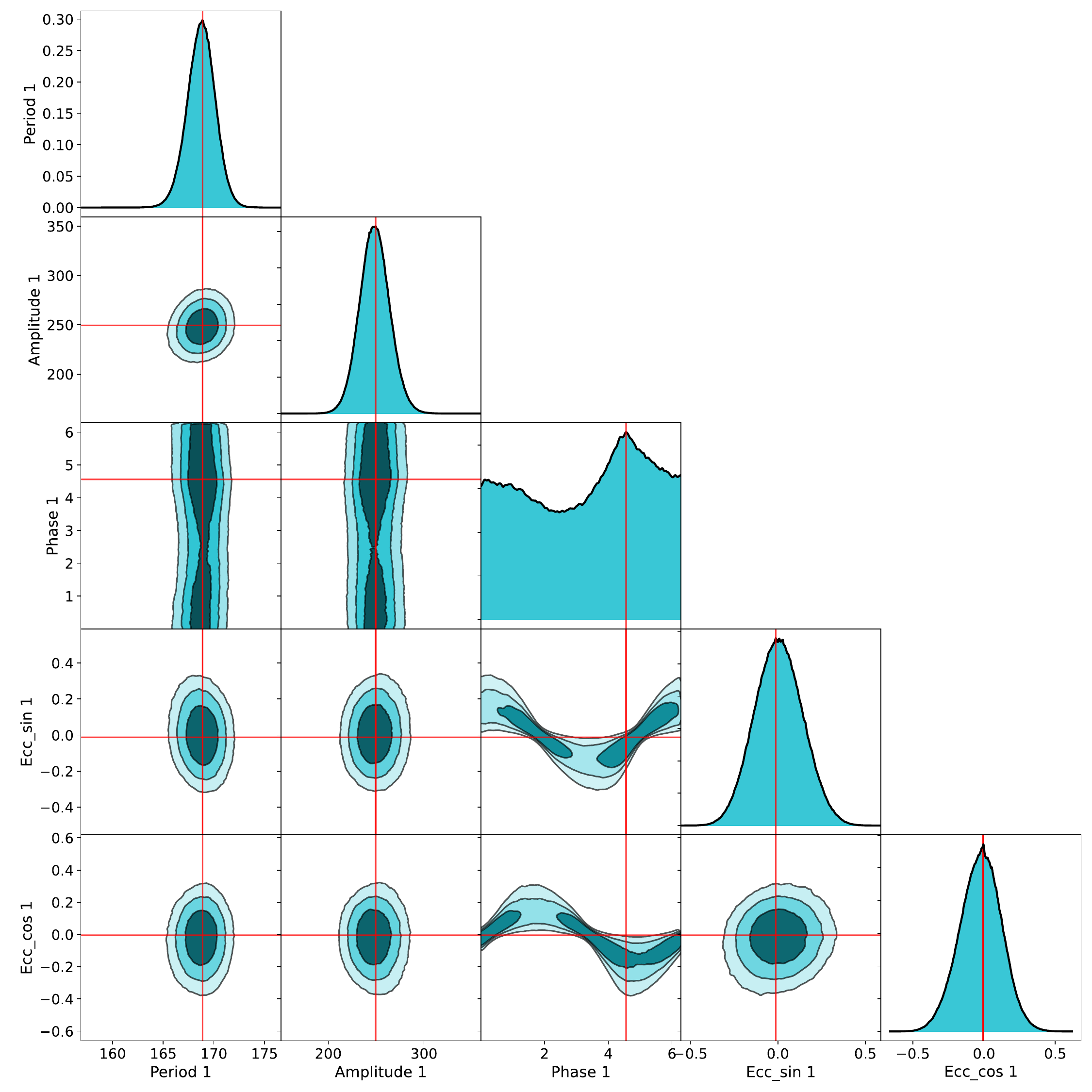}
    \caption{
        Corner plot for the orbital parameters of the larger, longer-period satellite.
    }
    \label{fig:extraperio1}
\end{figure}

\begin{figure}[t]
    \centering
    \includegraphics[width=\linewidth]{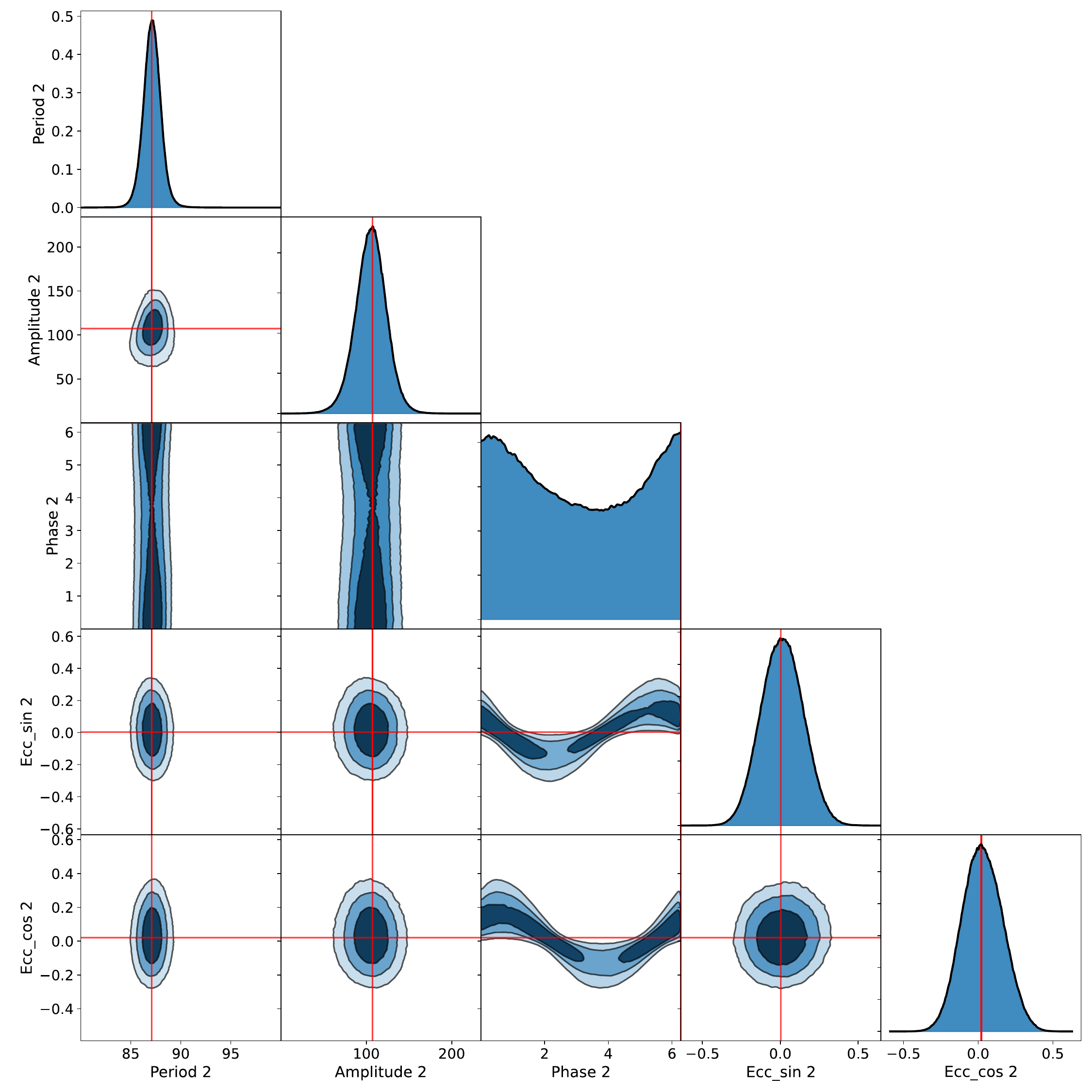}
    \caption{
        Corner plot for the orbital elements of the smaller, shorter-period satellite.
    }
    \label{fig:extraperio2}
\end{figure}
%\printbibliography

\clearpage
\setcounter{section}{0}

\section*{Author Contributions}

K.H. wrote the manuscript, analyzed the data, and obtained the results, A.Z. contributed to the writing of the article, conceived the idea, and led the survey. P.P. ran the \texttt{EMPEROR} analysis, J.K. assisted with the \texttt{viper} analysis, S.D., R.G., and C.L. worked on the observations and analysis. S.P. contributed target-specific insights and contextualization, F.R. and J.S. provided instrument-specific insights, V.d'O., I.C., and I.G. contributed to the observations execution. All the authors reviewed the article and provided feedback.

\section*{Competing Interests}

The authors declare no competing interests.

\section*{Materials \& Correspondence}

All correspondence and request for materials can be made to Kevin Hoy (kevin.hoy@mail.udp.cl).

\end{document}